\documentclass[twocolumn,prb,superscriptaddress,preprintnumbers,amsmath,amssymb,showpacs,floatfix]{revtex4}

\usepackage[dvips]{graphicx}
\usepackage{dcolumn}
\usepackage{bm}
\begin{document}

\title{Soft x-ray magnetic circular dichroism study on Gd-doped EuO thin films}

\author{H. Ott}
 \affiliation{ II. Physikalisches Institut, Universit\"{a}t zu K\"{o}ln,
 Z\"{u}lplicher Str. 77, 50937 K\"{o}ln, Germany}
\author{S. J. Heise}
 \affiliation{ II. Physikalisches Institut, Universit\"{a}t zu K\"{o}ln,
 Z\"{u}lplicher Str. 77, 50937 K\"{o}ln, Germany}
\author{R. Sutarto}
 \affiliation{ II. Physikalisches Institut, Universit\"{a}t zu K\"{o}ln,
 Z\"{u}lplicher Str. 77, 50937 K\"{o}ln, Germany}
\author{Z. Hu}
 \affiliation{ II. Physikalisches Institut, Universit\"{a}t zu K\"{o}ln,
 Z\"{u}lplicher Str. 77, 50937 K\"{o}ln, Germany}
\author{C. F. Chang}
 \affiliation{ II. Physikalisches Institut, Universit\"{a}t zu K\"{o}ln,
 Z\"{u}lplicher Str. 77, 50937 K\"{o}ln, Germany}
\author{H. H. Hsieh}
 \affiliation{Chung Cheng Institute of Technology,
 National Defense University, Taoyuan 335, Taiwan}
\author{H.-J. Lin}
 \affiliation{National Synchrotron Radiation Research Center,
 101 Hsin-Ann Road, Hsinchu 30077, Taiwan}
\author{C. T. Chen}
 \affiliation{National Synchrotron Radiation Research Center,
 101 Hsin-Ann Road, Hsinchu 30077, Taiwan}
\author{L. H. Tjeng}
 \affiliation{ II. Physikalisches Institut, Universit\"{a}t zu K\"{o}ln,
 Z\"{u}lplicher Str. 77, 50937 K\"{o}ln, Germany}

\date{\today}

\begin{abstract}
We report on the growth and characterization of ferromagnetic
Gd-doped EuO thin films. We prepared samples with Gd
concentrations up to 11\% by means of molecular beam epitaxy under
distillation conditions, which allows a very precise control of
the doping concentration and oxygen stoichiometry. Using soft
x-ray magnetic circular dichroism at the Eu and Gd $M_{4,5}$
edges, we found that the Curie temperature ranged from 69~K for
pure stoichiometric EuO to about 170~K for the film with the
optimal Gd doping of around 4\%. We also show that the Gd magnetic
moment couples ferromagnetically to that of Eu.
\end{abstract}


\maketitle

EuO is one of the rare ferromagnetic semiconductors and it has a
Curie temperature of 69~K.\cite{mauger86a} In slightly Eu-rich EuO
the low-temperature phase is metallic and the magnetic phase
transition is accompanied by a metal-to-insulator transition
(MIT), where the change in resistivity can exceed 10 orders of
magnitude depending on the exact
stoichiometry.\cite{oliver72a,shapira73a} An applied magnetic
field shifts the MIT temperature considerably which results in a
colossal magnetoresistance (CMR) effect with a resistivity change
of up to 8 orders of magnitude.\cite{shapira73a} The Curie
temperature $T_C$ can be strongly enhanced by electron doping. It
has been found that several percent of Gd doping enhances $T_C$ to
125~K, whereas oxygen deficiency can further increase $T_C$ to 160
K.\cite{mauger86a,matsumoto04a} In the ferromagnetic state the
unoccupied density of states shows a splitting of about 0.6~eV
between the spin-up and spin-down states leading to an almost
100\% spin polarization of the charge carriers in electron doped
EuO.\cite{steeneken02a} This essentially complete spin
polarization makes EuO a very attractive candidate for fundamental
research in the field of spintronics.

In this paper we show that Gd-doped EuO films can be prepared with
good control of the Gd concentration and oxygen stoichiometry
using the molecular beam epitaxy (MBE) technique under
distillation conditions. We investigated the doping dependence of
the ferromagnetic ordering temperature for a wide range of Gd
concentration. We used \textit{in-situ} soft x-ray absorption
spectroscopy (XAS) and magnetic circular dichroism (XMCD) in the
range of the Eu and Gd $M_{4,5}$ edges ($3d$$\rightarrow$$4f$) to
characterize the chemical composition and magnetic properties of
the films. The results are compared to predictions from mean-field
calculations. We also address the long standing issue whether the
Gd spins couple ferromagnetically or antiferromagnetically to the
Eu spins, and whether this coupling depends on the Gd
concentration.

The XAS and XMCD measurements were performed at the Dragon
beamline of the NSRRC in Taiwan using \textit{in-situ} MBE grown
samples. The spectra were recorded using the total electron yield
method in a chamber with a base pressure of 2$\times$10$^{-10}$
mbar. The photon energy resolution at the Eu and Gd $M_{4,5}$
edges ($h\nu \approx 1100$--1250~eV) was set at 0.6~eV, and the
degree of circular polarization was $\approx 80 \%$. For the XMCD
measurements the angle of x-ray incidence was set to $45^{\circ}$.
This angle is a compromise between a maximum projection of the
photon spin on the sample magnetization (the in-plane easy axis
requires grazing incidence) and the avoidance of saturation
effects, which may occur if the angle of incidence is too grazing,
because then the photon penetration depth becomes comparable to
the electron escape depth. The samples were placed in a magnetic
field of about 0.2~T using an \textit{ex-situ} rotatable permanent
magnet.

Gd-doped EuO films were grown \textit{in-situ} by MBE. High purity
Eu and Gd metal was evaporated from effusion cells and molecular
oxygen gas was supplied simultaneously through a leak valve. The
formation of oxides higher than EuO, such as Eu$_2$O$_3$ or
Eu$_3$O$_4$, as well as of Eu metal clusters has to be prevented.
This requires to set the Eu evaporation rate higher relative to
the oxygen dosing, such that the Eu is much in excess. At the same
time, the substrate temperature is kept at a sufficiently high
temperature so that a distillation process occurs in which excess
Eu which has not reacted to EuO is re-evaporated into the
vacuum.\cite{steeneken02b,tjeng05} The growth rate is in effect
determined by the oxygen dose rate. The use of the distillation
conditions is essential; otherwise one is confronted with the
difficult task to control very precisely the relative rates
between Eu and O to obtain (quasi) stoichiometric
EuO.\cite{matsumoto04a} The doping with Gd has been accomplished
by evaporating Gd and Eu metal simultaneously.

The Eu deposition rate was set at about 11~\AA/min, and the Gd
rate was varied between 0.1 and 2.7~\AA/min. The evaporation rate
was checked with a crystal thickness monitor. The oxygen partial
pressure was set at $6\times10^{-8}$ mbar above the base pressure
as monitored with a quadrupole mass spectrometer and kept constant
within $\pm0.2\times10^{-8}$ mbar. As substrates we used
epi-polished single crystal of Al$_2$O$_3$(1$\bar1$02) and
MgO(100). Prior to growth the substrates were annealed at
$T=600^{\circ}$C in the case of Al$_2$O$_3$ and $T=450^{\circ}$C
for MgO in an oxygen atmosphere of $1\times10^{-7}$ mbar in order
to obtain a clean and well-ordered substrate surfaces. The
substrates were kept at $T=350^{\circ}$C during growth.

\begin{figure}[t]
\includegraphics*[scale = 0.86] {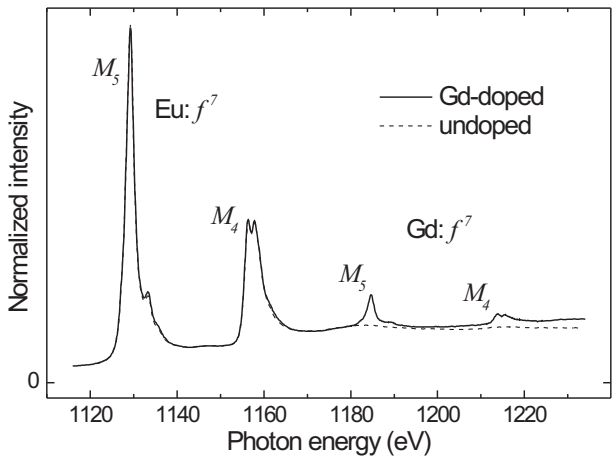}
\caption{\label{Fig_1:} Eu and Gd $M_{4,5}$
($3d$$\rightarrow$$4f$) XAS spectra of a stoichiometric (undoped)
EuO sample and a 9.9\% Gd-doped EuO sample.}
\end{figure}

Fig.~1 shows the XAS spectra of an undoped and a Gd-doped EuO film
across the Eu and Gd $M_{4,5}$ edges. The Eu spectra of the doped
and undoped case are identical and look very similar to the
theoretical spectrum calculated for a
$3d^{10}4f^{7}$$\rightarrow$$3d^{9}4f^{8}$
transition.\cite{thole85} This means that the Eu ions in our films
are divalent. Moreover, these experimental spectra look very
different from the one calculated for a Eu$^{3+}$
ion,\cite{thole85} and also have no extra peaks at higher energies
which otherwise would indicate the presence of Eu$^{3+}$
ions.\cite{holroyd04} All this demonstrates that our EuO films are
indeed free from Eu$^{3+}$ contamination.

The Gd spectrum in Fig.~1 has also all the characteristics of a
$3d^{10}4f^{7}$$\rightarrow$$3d^{9}4f^{8}$
transition.\cite{thole85} This is consistent with the fact that Gd
has always the $4f^{7}$ configuration. The fact that Eu and Gd
have the identical $4f$ configuration and very similar spectral
line shapes facilitates the determination of the Gd concentration
in the doped EuO films: we can simply deduce this from the ratio
of the main-peak heights of the Gd and Eu spectra after
subtracting the extended x-ray absorption fine structure (EXAFS)
of pure EuO in the Gd $M_{4,5}$ energy range. This is a simple and
reliable procedure with the advantage that the Gd concentration
can be determined \textit{in-situ}. For the particular Gd-doped
film shown in Fig.~1 we find that the Gd concentration is about
9.9\%.

\begin{figure}[t]
\includegraphics*[scale = 0.52] {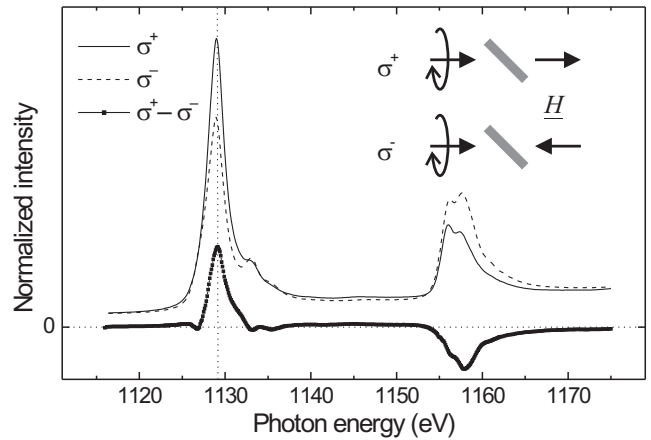}
\caption{\label{Fig_2:} Eu $M_{4,5}$ XAS spectra of a 3.7\%
Gd-doped EuO sample recorded at 20 K using circularly polarized
x-rays with the photon spin parallel (solid line) and antiparallel
(dashed line) to the magnetic field direction. The lowest curve
shows the difference between both, which is called the XMCD
spectrum.}
\end{figure}

The magnetic properties of the samples have been investigated by
XMCD.\cite{goedkoop88a} Fig.~2 shows the Eu $M_{4,5}$ spectra of a
3.7\% Gd-doped EuO sample recorded at a temperature of 20~K using
circularly polarized x-rays with the photon spin parallel and
antiparallel to the magnetic field direction. Clearly the two
spectra show significant differences; the difference spectrum,
i.e. the XMCD spectrum, is given by the lowest curve. The largest
XMCD signals can be observed at 1129.1~eV, which is about 0.1~eV
higher in energy than the maximum of the $M_5$ white line, and at
1157.9~eV for the $M_4$. This XMCD spectrum matches very nicely
the theoretical spectrum for a Eu$^{2+}$ ion.\cite{goedkoop88a}
The experimental XMCD effect, defined as the difference divided by
the sum, becomes 28\% and $-$41\% for the $M_5$ and the $M_4$
white line, respectively. In these numbers the angle of incidence
and the degree of photon polarization have been taken into
account. Since about 51\% XMCD effect is expected theoretically at
the $M_5$ white line for a fully magnetized Eu$^{2+}$
ion,\cite{goedkoop88a} we conclude that the degree of
magnetization in this measurement is about 55\%.

\begin{figure}[t]
\includegraphics*[scale = 0.67] {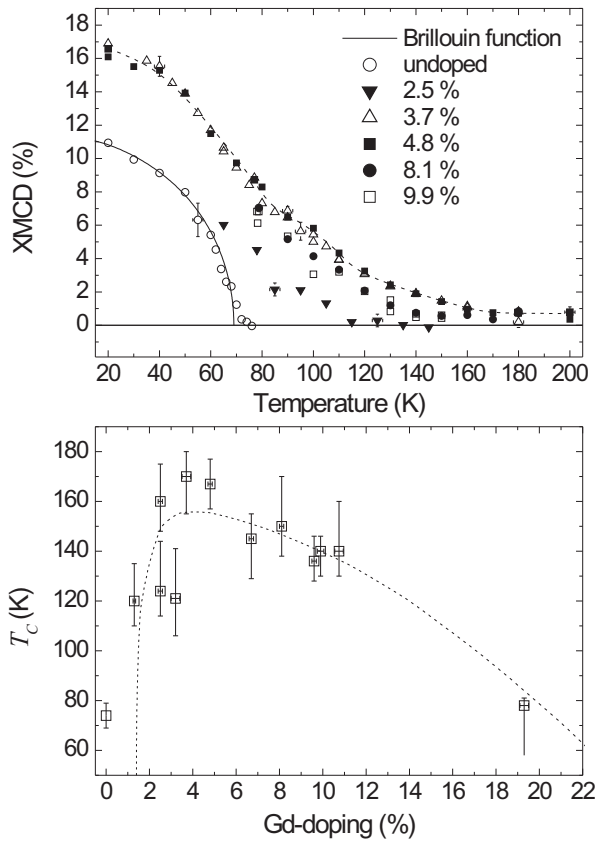}
\caption{\label{Fig_3:} Top panel: XMCD effect as a function of
temperature for EuO samples with different Gd doping levels.
Bottom panel: Doping dependence of the Curie temperature ($T_C$).
The dotted line represents the result of a mean-field
calculation.}
\end{figure}

The top panel of Fig.~3 depicts the XMCD effect in pure EuO and
several Gd-doped EuO thin films with different doping levels as a
function of temperature. The XMCD effect has been evaluated at the
photon energy of 1129.1~eV. The undoped EuO sample (open circles)
clearly follows a Brillouin function with a $T_C$ of about 69 K,
identical to the value for bulk EuO. However, upon doping with Gd
the magnetization increases, the shape of the temperature
dependence of the XMCD effect deviates strongly from the Brillouin
function, and $T_C$ is enhanced considerably. Here we took as
$T_C$ the temperature above which the XMCD signal is constant. The
bottom panel of Fig.~3 shows $T_C$ as a function of the Gd doping
concentration. Starting from around 69~K for undoped EuO, $T_C$
increases rapidly upon Gd doping and reaches a maximum of about
170~K at a doping concentration of about 4\%. To our knowledge,
this is the highest $T_C$ reported so far for a EuO system under
ambient pressures. For higher Gd concentrations $T_C$ slowly
decreases again.

The deviation from the Brillouin function of the temperature
dependence of the XMCD effect is qualitatively in agreement with
magnetization curves reported earlier for bulk samples of Gd-doped
EuO. Mauger explains this by the temperature dependence of the
effective magnetic coupling due to the successive population of
the spin-up and spin-down conduction subband which enters the
exchange coupling constant.\cite{mauger77a} The doping dependence
of $T_C$ has also been modelled by Mauger.\cite{mauger77a} The
bottom panel of Fig.~3 shows the reproduced curve calculated using
a mean-field approximation. The calculation assumes a critical Gd
concentration of about 1\%. Below this concentration it is
energetically favorable for the ``extra'' electrons to remain
localized around the Gd impurities, forming a bound magnetic
polaron. Consequently there is no indirect exchange mediated via
free carriers, and $T_C$ is not enhanced with respect to undoped
EuO. Above the critical Gd concentration, free carriers are
available and $T_C$ increases accordingly. There is a maximum in
$T_C$ which according to the model is due to the instability of
the ferromagnetic configuration with respect to a spiral
configuration along the [111] direction when the concentration is
too high.\cite{mauger77a} The agreement between experiment and
theory appears to be satisfactory.

An open question so far is whether the Gd spins in Gd-doped EuO
are coupled to the Eu spins, and if so, whether they are parallel
or antiparallel aligned. Until now little has been reported on
this subject in literature. From the few studies carried out in
the past it was suggested that the Gd spins may be aligned
antiparallel to the Eu spins in Gd-doped EuS
films~\cite{mcguire71a} and single crystals.\cite{bayer71a} Now,
with the XMCD technique being developed into maturity over the
last 15 years, the issue of spin alignment can be addressed in a
straightforward manner.

\begin{figure}[t]
\includegraphics*[scale = 1.1] {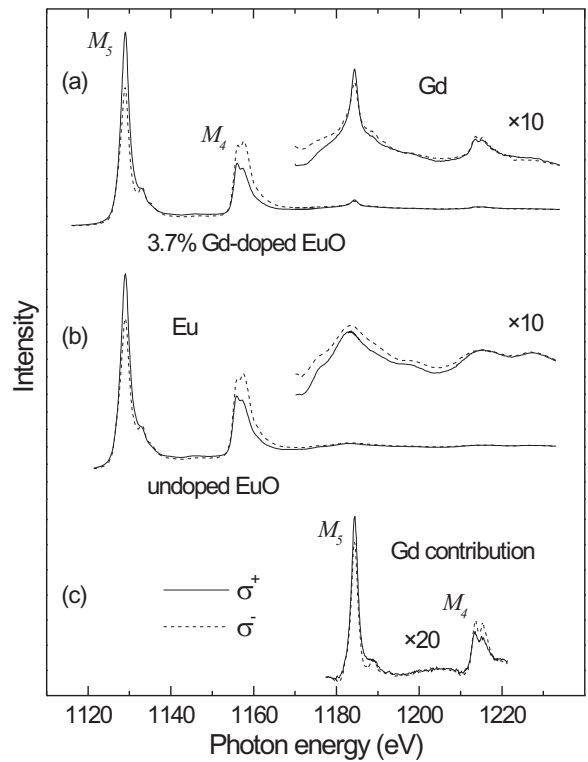}
\caption{\label{Fig_4:} Eu and Gd $M_{4,5}$ XAS spectra of (a) a
3.7\% Gd-doped EuO sample and (b) an undoped EuO sample, recorded
at 20~K using circularly polarized x-rays with the photon spin
parallel and antiparallel to the magnetic field direction. (c) The
net Gd $M_{4,5}$ contribution, obtained by subtracting the EuO
spectra from those of the 3.7\% Gd-doped EuO.}
\end{figure}

In panel (a) of Fig.~4 we present the Eu and Gd $M_{4,5}$ spectra
of a 3.7\% Gd-doped EuO film taken at 20 K using circularly
polarized light. The XMCD effect in the Eu $M_{4,5}$ edges can be
clearly seen, with the $M_5$ peak having larger intensity for
$\sigma^{+}$ polarized x-rays (solid line) than for $\sigma^{-}$
polarization (dashed line). Since the intensity of the Gd
contribution is relatively weak, we have magnified the Gd part of
the spectra by a factor of ten. The XMCD effect for the Gd edges
is now clearly visible, but we also observe that the background
changes with the polarization of the light. We attribute this to
the presence of the extended x-ray absorption fine structure
(EXAFS) of the Eu edges, which is superimposed on the Gd spectra.
As can be seen in panel (b) of Fig.~4, the Eu EXAFS of undoped EuO
indeed carries an XMCD effect in the photon energy region of the
Gd edge. To resolve this background problem, we subtract the
spectrum of the undoped EuO from that of the Gd-doped EuO. The
resulting net Gd contribution to the spectra is shown in panel (c)
of Fig.~4, and the similarity of the Gd spectra with those of Eu
is striking. In particular, the Gd $M_5$ peak with $\sigma^+$
light lies above that with $\sigma^-$, i.e. identical to the Eu
case. This directly means that the Gd and Eu $4f$ spins are
aligned parallel. We have investigated the Gd $4f$ spin alignment
for various doping levels up to 11\%, and found in all cases that
it is parallel to the Eu spins, in agreement with the analysis of
Mauger.\cite{mauger77a}

In conclusion, we have successfully prepared Gd-doped EuO films
with Gd concentrations up to 11\% with controlled stoichiometry by
means of molecular beam epitaxy. The magnetic ordering temperature
is enhanced upon Gd doping and a record high $T_C\approx 170$~K
has been achieved for an optimal Gd concentration of around 4\%.
We also revealed that the Gd magnetic moments couple
ferromagnetically to the magnetic moments of Eu.

We acknowledge the NSRRC staff for providing us with an extremely
stable beam. We would like to thank Lucie Hamdan for her skillful
technical and organizational assistance in preparing the
experiment. The research in Cologne is supported by the Deutsche
Forschungsgemeinschaft through SPP 1166 and SFB 608.


\begin{references}


\bibitem{mauger86a}
For a review, see A.~Mauger and C.~Godart, Phys. Rep.
\textbf{141}, 51 (1986).

\bibitem{oliver72a}
M. R. Oliver, J. O. Dimmock, A. L. McWhorter, and T. B. Reed,
Phys. Rev. B \textbf{5}, 1078 (1972).

\bibitem{shapira73a}
Y. Shapira, S. Foner, R. L. Aggarwal, and T. B. Reed, Phys. Rev. B
\textbf{8}, 2299; \textbf{8}, 2316 (1973).

\bibitem{matsumoto04a}
T.~Matsumoto, K.~Yamaguchi, M.~Yuri, K.~Kawaguchi, N.~Koshizaki,
and K.~Yamada, J. Phys.: Condens. Matter \textbf{16}, 6017 (2004).

\bibitem{steeneken02a}
P.~G. Steeneken, L.~H. Tjeng, I.~Elfimov, G.~A. Sawatzky,
G.~Ghiringhelli, N.~B. Brookes, and D.-J. Huang, Phys. Rev. Lett.
\textbf{88}, 047201 (2002).

\bibitem{steeneken02b}
P.~G.~Steeneken, Ph.D. thesis, Groningen (2002).

\bibitem{tjeng05}
L.~H.~Tjeng \textit{et al.}, in preparation.

\bibitem{thole85} B. T. Thole, G. van~der~Laan, J. C. Fuggle, G. A. Sawatzky,
R. C. Karnatak, and J.-M. Esteva, Phys. Rev. B \textbf{32}, 5107
(1985).

\bibitem{holroyd04} J. Holroyd, Y. U. Idzerda, and S. Stadler,
J. Appl. Phys. \textbf{95}, 6571 (2004).

\bibitem{goedkoop88a}
J. B. Goedkoop, B. T. Thole, G. van~der~Laan, G. A. Sawatzky, F.
M. F. de~Groot, and J. C. Fuggle, Phys. Rev. B \textbf{37}, 2086
(1988).

\bibitem{mauger77a}
A.~Mauger, Phys. Stat. Sol. \textbf{84}, 761 (1977).

\bibitem{mcguire71a}
T.~R. Mc{G}uire and F.~Holtzberg, in: \textit{Magnetism and
Magnetic Materials}, \textit{AIP Conf. Proc. No. 5}, vol.~2,
855--859 (AIP, 1971).

\bibitem{bayer71a}
E.~Bayer and W.~Zinn, Z. Angew. Phys. \textbf{32}, 83 (1971).

\end{references}
\end{document}